\documentclass[twocolumn,showpacs]{revtex4}
\begin{document}
\title{Energy conservation for dynamical black holes}
\author{Sean A. Hayward}
\affiliation{Department of Science Education, Ewha Womans University,
Seodaemun-gu, Seoul 120-750, Korea\\ {\tt sean\_a\_hayward@yahoo.co.uk}}
\date{revised 3rd August 2004}

\begin{abstract}
An energy conservation law is described, expressing the increase in mass-energy
of a general black hole in terms of the energy densities of the infalling
matter and gravitational radiation. For a growing black hole, this first law of
black-hole dynamics is equivalent to an equation of Ashtekar \& Krishnan, but
the new integral and differential forms are regular in the limit where the
black hole ceases to grow. An effective gravitational-radiation energy tensor
is obtained, providing measures of both ingoing and outgoing, transverse and
longitudinal gravitational radiation on and near a black hole. Corresponding
energy-tensor forms of the first law involve a preferred time vector which
plays the role for dynamical black holes which the stationary Killing vector
plays for stationary black holes. Identifying an energy flux, vanishing if and
only if the horizon is null, allows a division into energy-supply and work
terms, as in the first law of thermodynamics. The energy supply can be
expressed in terms of area increase and a newly defined surface gravity,
yielding a Gibbs-like equation, with a similar form to the so-called first law
for stationary black holes.
\end{abstract}
 \maketitle

{\em Introduction.} Ashtekar \& Krishnan \cite{AK1,AK2,Ash} recently derived an
energy-balance equation for dynamical black holes, expressing the increase in
mass-energy of a black hole in terms of the energy densities of the infalling
matter and gravitational radiation. This was an important step in ongoing
efforts to develop a theory of black-hole dynamics, as opposed to the textbook
theory of black holes \cite{BCH,HE,MTW,Wal}, which mostly concerns either
stationary space-times or event horizons, which cannot be located by mortals.
The Ashtekar-Krishnan equation describes how a black hole grows, but does not
include the physically important limit where it is starved of nourishment and
ceases to grow. This Letter summarizes some key results of an approach which
can be applied to black holes in any state. In particular, an
energy-conservation equation is found in several different forms, including a
generalization of the Ashtekar-Krishnan equation. Full details are presented in
a longer article \cite{bhd3}.

The key geometrical objects are trapping horizons \cite{bhd,bhs}, which are
hypersurfaces $H$ (in space-time) foliated by marginal surfaces. A marginal
surface is a spatial surface on which one null expansion vanishes, where the
null expansions $\theta_\pm$ may be defined by $\theta_\pm d^2A=L_\pm(d^2A)$,
where $d^2A$ denotes the area element and $L_\pm$ the Lie derivatives along
future-pointing null normal vectors $l_\pm$. This is a surface where outgoing
or ingoing light rays are just trapped, neither converging nor diverging. For
the author's local definition of black hole as a future ($\theta_-<0$ on $H$,
for $\theta_+=0$) outer ($L_-\theta_+<0$) trapping horizon, it was shown that,
assuming the dominant energy condition, the marginal surfaces have spherical
topology (if compact) and that $H$ is either spatial or null at each point,
with its rate of area increase being respectively positive or zero \cite{bhd}.
This is similar to Hawking's so-called second law for event horizons, but for a
physically locatable horizon.

In this Letter, the new results apply to any trapping horizon, and can
therefore be applied to black holes, white holes, traversible wormholes
\cite{wh} and cosmological horizons. In comparison, the Ashtekar-Krishnan
formalism applies only to spatial trapping horizons. The underlying aim of this
framework for black-hole dynamics is to provide a local, physical understanding
of black holes, which can be used to guide and interpret more detailed studies,
including numerical computations, and so help to make better contact with
astronomical observations.

{\em Method.} Einstein gravity is assumed, with space-time metric $g$.
Labelling the marginal surfaces by a coordinate $x$, $H$ is generated by a
vector $\xi=\partial/\partial x$, normal to the marginal surfaces. The
coordinate freedom here is just $x\mapsto\hat x(x)$ and choice of angular
coordinates, under which all the key formulas will be invariant. To study
derivatives off $H$, one may construct a dual-null foliation \cite{dn,dne}: two
families of null hypersurfaces, forming the wavefronts of ingoing and outgoing
radiation, intersecting in the marginal surfaces. Such a dual-null foliation
always exists locally, and for spatial $H$, it is unique. For null $H$, it is
not unique, leaving some subtleties to be resolved. Labelling the null
hypersurfaces by dual-null coordinates $x^\pm$, one has normal 1-forms
$n^\pm=-dx^\pm$ such that $g^{-1}(n^\pm,n^\pm)=0$, a normalization function
$g^{+-}=g^{-1}(n^+,n^-)<0$, and $l_\pm=-g^{-1}(n^\mp)/g^{+-}$ are the normal
projections of the evolution vectors $\partial/\partial x^\pm$. One can fix
$g^{+-}\cong-1$, where $\cong$ denotes evaluation on $H$, but it will be
retained here, since it cannot generally be so fixed away from $H$.

The area of the transverse surfaces $S$ is $A=\oint_Sd^2A$ and the area radius
$R=\sqrt{A/4\pi}$ is convenient. As a measure of the active gravitational
mass-energy enclosed by a surface, it is useful to take the Hawking energy
\cite{Haw}
\begin{equation}\label{energy}
E=\frac R{2G}\left(1-\frac1{8\pi}\oint_Sg^{+-}\theta_+\theta_-d^2A\right)
\end{equation}
where $c=1$ units are adapted to the dual-null method.

{\em First law.} The first result is an energy conservation law which will here
be called the first law of black-hole dynamics:
\begin{equation}\label{first}
\frac{\partial E}{\partial
x}\cong\oint_S\xi\cdot\psi\,d^2A+\oint_Sw\,d^2A\,\frac{\partial R}{\partial x}
\end{equation}
where $(\psi,w)$ are defined below. The equation follows from the Einstein
equation, specifically the null focusing and cross-focusing equations
\cite{bhd} for $(L_\pm\theta_\pm,L_\mp\theta_\pm)$, employing the Gauss-Bonnet
and Gauss divergence theorems on $S$. The Ashtekar-Krishnan derivation has some
similarities, but uses a different decomposition of the Einstein equation,
based on the conventional 3+1 formalism for spatial $H$. Dual-null foliations
are generally less familiar, but much better adapted to radiation, which
propagates outwards in advanced time $x^+$ as a profile in retarded time $x^-$
at given angle.

The 1-form $\psi$ is an energy flux (or the dual vector is an energy-momentum
density) and the function $w$ is an energy density, and they are each composed
of matter and gravitational parts, $\psi=\psi_m+\psi_g$, $w=w_m+w_g$. The
matter parts are defined in terms of the matter energy tensor $T$ as
\begin{eqnarray}
w_m&=&-\hbox{trace}\,T/2\\
\psi_m&=&T\cdot\nabla R+w_m\nabla R
\end{eqnarray}
where $\nabla$ is the covariant derivative and the trace is in the space normal
to $S$. The expressions have the same form as those in spherical symmetry
\cite{1st}, in which case the gravitational terms vanish. The coordinate forms
are
\begin{eqnarray}
w_m&=&-g^{+-}T_{+-}\\
(\psi_m)_\pm&=&g^{+-}T_{\pm\pm}L_\mp R.
\end{eqnarray}
As simple examples in spherical symmetry, for a massless Klein-Gordon field
$\phi$, $T_{\pm\pm}=(\partial_\pm\phi)^2$ are the energy densities of radiation
propagating in the $l_\mp$ directions, with $T_{+-}=0$, whereas one finds the
electric energy density $w_m={\cal E}^2/8\pi$ for an electric field ${\cal
E}=q/R^2$, where $q$ is the enclosed charge \cite{1st}. Then one can interpret
$\oint_S\xi\cdot\psi_md^2A$ as a rate of energy supply and $\oint_Sw_md^2A$ as
a rate of work.

The gravitational energy flux is found to have components
\begin{equation}
(\psi_g)_\pm=g^{+-}||\sigma_\pm||^2L_\mp R/32\pi G
\end{equation}
where the traceless bilinear forms $\sigma_\pm=\bot L_\pm h-\theta_\pm h$ are
the null shear tensors, $h$ denotes the metric of $S$ and $\bot$ projection by
$h$. The expressions for $\psi_g$ have a similar form to expressions for the
energy flux of transverse gravitational radiation in several limits where it is
well defined: at infinity in an asymptotically flat space-time \cite{mon,inf},
for linearized gravitational radiation in the shortwave approximation
\cite{MTW}, in cylindrical symmetry \cite{cyl} and in a quasi-spherical
approximation \cite{qs,gwbh,gwe}. Indeed, $\psi$ reduces to the Bondi flux at
null infinity. Thus one may physically interpret $\psi_g$ as energy flux of
transverse gravitational radiation and $\oint_S\xi\cdot\psi\,d^2A$ as the rate
of energy supply to the black hole by the infalling matter and gravitational
radiation. Finally one has
\begin{equation}
w_g=|\zeta|^2/8\pi G
\end{equation}
where the 1-forms $\zeta_{(\pm)}=-\bot((n^\mp\cdot\nabla)n^\pm)/g^{+-}$ are
normal fundamental forms of $S$, and one uses $\zeta=\zeta_{(\pm)}$ for $H$
with $\theta_\pm\cong0$. For reasons best seen in a spin-coefficient
formulation, $w_g$ can be interpreted as the energy density of ingoing
longitudinal gravitational radiation \cite{bhd4}.

In summary, the first law (\ref{first}) expresses the rate of change of
black-hole mass as a sum of two terms, interpreted respectively as rate of
energy supply and rate of work. The terminology is analogous to that of the
first law of thermodynamics, which expresses the change of internal energy
(heat) as a sum of heat supply and work. Heat flux vanishes in thermal
equilibrium, and a similar property holds for black holes: the energy flux
$\xi\cdot\psi$ vanishes if $H$ is null, whereas the work density $w$ is
generally non-zero for $H$ of any causal nature. This is a physically important
distinction because (as mentioned) $H$ is spatial for a growing black hole, but
becomes null in the limit where it ceases to grow and reaches equilibrium.

{\em Integral forms.} When the distinction between energy-supply and work terms
is unimportant, the first law may be written simply as
\begin{equation}
\frac{\partial E}{\partial x}\cong\oint_S\epsilon\frac{\partial R}{\partial
x}d^2A
\end{equation}
where $\epsilon=w+\xi\cdot\psi/(\partial R/\partial x)$ is the combined energy
density, which can be divided into that due to the matter and the gravitational
field in the obvious way, $\epsilon=\epsilon_m+\epsilon_g$. A corresponding
integral form of the first law is
\begin{equation}
[E]\cong\int_H\epsilon\frac{\partial R}{\partial x}d^2A\,dx
\end{equation}
where $[E]$ denotes the change in $E$ along the horizon, from one marginal
surface to another. This expression uses the generator-volume element
$d^2A\,dx$, which is regular as $H$ becomes null. Ashtekar \& Krishnan used the
proper-volume element $d^3v=\sqrt{g_{xx}}d^2A\,dx$, where $g_{xx}=g(\xi,\xi)$,
which yields the proper-volume form
\begin{equation}
[E]\cong\int_H\tilde\epsilon\,d^3v
\end{equation}
where $\tilde\epsilon=\epsilon(\partial R/\partial x)/\sqrt{g_{xx}}$ is the
proper energy density. This form is technically regular in the limit where $H$
becomes null, where $g_{xx}\to0$, but less useful than the differential or
generator-volume forms, since the proper-volume element vanishes, $d^3v\to0$,
with $\tilde\epsilon$ being generally finite. Conversely, the integrand
$\epsilon(\partial R/\partial x)$ of the generator-volume form vanishes in the
null limit.

{\em Energy-tensor forms.} The horizon induces a preferred time vector $\chi$
which is the curl of $R$ in the normal space: $\chi$ is orthogonal to $R$ and
the transverse surfaces, $\chi\cdot dR=0$, $\bot\chi=0$, has normalization
$g(\chi,\chi)=-g^{-1}(dR,dR)$ and becomes null on a trapping horizon,
$g(\chi,\chi)\cong0$. This generalizes the Kodama vector in spherical symmetry
\cite{1st,sph}. In particular, $\chi$ reduces to the stationary Killing vector
for Schwarzschild and Reissner-Nordstr\"om black holes, so it can be regarded
as playing the traditional role of a stationary Killing vector even for
dynamical black holes.

One can also introduce the vector $\tau$ which is dual to the
horizon-generating vector $\xi$ in the normal space: $\tau$ is normal to $H$,
$g(\xi,\tau)=0$, $\bot\tau=0$, has normalization $g(\tau,\tau)=-g(\xi,\xi)$ and
is regular in the null limit, becoming null itself, $\tau\to\xi$. Then the
matter energy density is found to have a remarkably simple form:
\begin{equation}
\epsilon_m\frac{\partial R}{\partial x}=T(\chi,\tau).
\end{equation}
In spherical symmetry, one actually has $\partial E/\partial x=AT(\chi,\tau)$
for any normal vector $\chi$ and its orthogonal dual $\tau$. Dividing the
integrated flux as $[E]=[E]_m+[E]_g$, it follows that the integrated matter
flux is
\begin{equation}
[E]_m\cong\int_HT(\chi,\tau)d^2A\,dx.
\end{equation}
This is reminiscent of the usual definition of energy in a stationary
space-time as a volume integral, with $\chi$ replacing the stationary Killing
vector. Ashtekar \& Krishnan gave a similar form, which here would be
\begin{equation}
[E]_m\cong\int_HT(\chi,\hat\tau)d^3v
\end{equation}
where $\hat\tau=\tau/\sqrt{g_{xx}}$ is the unit normal vector to $H$. This
works for spatial trapping horizons, but not in the null limit, where a unit
normal vector does not exist. However, it reveals that the Ashtekar-Krishnan
permissible vector fields are gauge-fixed versions (with $g^{+-}\cong-1$,
$\xi^+\cong-\xi^-$) of the more manifestly invariant $\chi/(\partial R/\partial
x)$. The Killing-like vector $\chi$ seems to have fundamental importance for
dynamical black holes, defining a preferred flow of time outside the black
hole.

The integrated flux due to gravitational terms can be written in a similar way
by introducing an effective gravitational-radiation energy tensor $\Theta$ in
the normal space, with components
\begin{eqnarray}
\Theta_{\pm\pm}&=&||\sigma_\pm||^2/32\pi G\\
\Theta_{\pm\mp}&=&-|\zeta_{(\pm)}|^2/8\pi Gg^{+-}.
\end{eqnarray}
Since these components are non-negative, $\Theta$ satisfies the dominant energy
condition, implying that the gravitational radiation carries positive energy.
The generally non-symmetric nature of $\Theta$ is curious, but does give it the
correct number of components for the energy-momentum density of ingoing and
outgoing, transverse and longitudinal gravitational radiation. Note also that
$(\sigma_\pm,\zeta_{(\pm)})$ each have the correct number (two) of independent
components for describing the respective radiation. The identification and neat
division of these modes is another success for the dual-null method.

Then the differential form of the first law is
\begin{equation} \frac{\partial
E}{\partial x}\cong\oint_S(T(\chi,\tau)+\Theta(\chi,\tau))d^2A
\end{equation}
and the generator-volume form is
\begin{equation}
[E]\cong\int_H(T(\chi,\tau)+\Theta(\chi,\tau))d^2A\,dx.
\end{equation}
If the trapping horizon is spatial, one can also write
\begin{equation}
[E]\cong\int_H(T(\chi,\hat\tau)+\Theta(\chi,\hat\tau))d^3v
\end{equation}
which is closest to the Ashtekar-Krishnan form, having identified $\chi$ and
$\Theta$. These three forms perhaps best illustrate the nature of the first law
as an energy-conservation equation, expressing the increase in the mass-energy
$E$ of the black hole due to the energy densities of the infalling matter and
gravitational radiation.

{\em Gibbs-like form.} The so-called first law for stationary black holes
\cite{BCH,HE,MTW,Wal} is actually analogous to the Gibbs equation rather than
the first law of thermodynamics; both are assumed independently in
non-equilibrium thermodynamics, formulated as a local field theory
\cite{Eck,dGM,MR,th}. To find such an equation for dynamical black holes, one
needs a definition of surface gravity $\kappa$ which generalizes the standard
definition for stationary black holes. In spherical symmetry, there is a
definition which has several desired properties \cite{1st,in}, including a
relation $\chi\cdot(\nabla\wedge\chi)\cong\pm\kappa\chi$ with the same form as
the standard one. A simple generalization, modifying an earlier definition
\cite{MH}, is
\begin{equation}\label{sg}
\kappa=g^{+-}\left(2L_-\theta_++\theta_+\theta_-\right)R/4
\end{equation}
for $\theta_+\cong0$. This has the desired property that it vanishes where
$L_-\theta_+\cong0$, which is the previously defined condition for degenerate
trapping horizons \cite{bhd}. Then the rate of energy supply satisfies
\begin{equation}
\oint_S\xi\cdot\psi\,d^2A\cong\frac{1}{8\pi GA}\oint_S\kappa\frac{\partial
A}{\partial x}d^2A.
\end{equation}
This is a remarkable generalization of the $\kappa\delta A/8\pi G$ term in the
so-called first law for stationary black holes, or the $\kappa(\partial
A/\partial x)/8\pi G$ term found previously in spherical symmetry \cite{1st}.
Thus the desired Gibbs-like equation is found simply as
\begin{equation}\label{gibbs}
\frac{\partial E}{\partial x}\cong\frac{1}{8\pi GA}\oint_S\kappa\frac{\partial
A}{\partial x}d^2A+\oint_Sw\,d^2A\,\frac{\partial R}{\partial x}.
\end{equation}
Since stationary black holes have a Hawking temperature $\kappa\hbar/2\pi k$
and an inferred entropy $S\cong Ak/4\hbar G$, the above result suggests that
the same is true locally for dynamical black holes, by analogy with the
original Clausius concept of entropy, as previously argued in spherical
symmetry \cite{HMA}. Defining a geometric entropy flux vector $\varphi=2\pi\psi
k/\hbar\kappa$ and a geometric entropy supply $\partial S_\circ/\partial
x=\oint_S\xi\cdot\varphi\,d^2A$, a geometric entropy conservation equation
$\partial S/\partial x\cong\partial S_\circ/\partial x$ can also be obtained
\cite{bhd3}.

When comparing $\kappa$ with the standard definition of surface gravity for
stationary black holes, one faces the problem that the dual-null foliation is
generally not unique for null $H$, and therefore neither is $\chi$ nor
$\kappa$. It becomes unique in spherical symmetry, where $\kappa$ reduces to
the standard surface gravity $1/4Gm$ for Schwarzschild and, non-trivially,
$\sqrt{m^2-q^2}/G(m+\sqrt{m^2-q^2})^2$ for Reissner-Nordstr\"om black holes
\cite{1st}. For Kerr black holes, a dual-null foliation giving the correct
$\chi$ (stationary Killing vector) and $\kappa$ should exist, but is not known
explicitly, despite a recent effort \cite{kerr}. It should also be noted that
the definition (\ref{sg}) of surface gravity is inequivalent (for spatial
horizons) to that of Ashtekar \& Krishnan, and that their generalization of the
so-called first law also has a different form to the Gibbs-like equation
(\ref{gibbs}). Similar remarks apply to the approach of Booth \& Fairhurst
\cite{BF}.

{\em Conclusion.} The most practical result here is probably the identification
of the effective gravitational-radiation energy tensor $\Theta$. A new
generation of gravitational-wave detectors offer an entirely new window on the
universe, and a considerable community is attempting to produce predictions of
waveforms from cataclysmic events such as black-hole mergers, mainly by
numerical simulations. Classically, gravitational radiation is well defined
only in weak-field regimes, which may be outside the numerical domain; the
radiation-extraction problem. However, the new $\Theta$ is defined on a
black-hole horizon $H$ (as located by existing numerical methods) and in its
vicinity, and also yields the standard Bondi flux of gravitational radiation at
null infinity. Thus a computational module could lock on to $H$, transform the
raw numerical data from the original coordinates to the dual-null foliation and
display the conformal energy flux $R^2\psi_-$ of the outgoing radiation at a
given angle, as a graph against retarded time $x^-$, running as video in
advanced time $x^+$, and so provide a graphic view of whether the energy
profiles are converging in advanced time. One could also find the conformal
strain tensor $\varepsilon=\int(\sigma_-/2R)dx^-$, which gives the strain
tensor $\varepsilon/R$ to be measured by a gravitational-wave detector at large
distance $R$ \cite{gwbh}. Remarkably, gravitational radiation appears to be
well defined in the strong-field regime, the dynamical black hole itself
providing the required structure.

The discovery of a comparatively simple energy conservation law for completely
general black holes is itself remarkable, as emphasized by Ashtekar \cite{Ash}.
The black hole might be a recent black-hole merger, violently distorted and
gorging voraciously on ambient matter and radiation, yet the law dictates
precisely how its mass and area grow.  A quite common view has been that the
basic principles of black-hole mechanics were understood thirty years ago
\cite{BCH,HE,MTW} but that the complexity of the Einstein equations precludes a
detailed physical understanding of the dynamical evolution of black holes. A
new picture is emerging, in which black holes grow and evolve with geometrical
elegance, obeying simple physical laws.

\smallskip\noindent This work was inspired by a presentation of Abhay Ashtekar
at the Erwin Schr\"odinger Institute in Vienna. The author thanks the ESI for
local support and hospitality, and attendees for discussions.

\end{document}